\documentclass{article}
\usepackage{graphicx} 
\usepackage{amsmath}
\usepackage{hyperref}
\usepackage{authblk}
\usepackage{tcolorbox}
\usepackage{enumitem}
\usepackage{cite}
\title{Impermanent loss and loss-vs-rebalancing I:\\ some statistical properties}

\author{
  Abe Alexander\thanks{abealexander@outlook.com} \and
  Lars Fritz\thanks{lsfritz@proton.me}
}

\begin{document}

\maketitle
\begin{abstract}
There are two predominant metrics to assess the performance of automated market makers and their profitability for liquidity providers: 'impermanent loss' (IL) and 'loss-versus-rebalance' (LVR). In this short paper we shed light on the statistical aspects of both concepts and show that they are more similar than conventionally appreciated. Our analysis uses the properties of a random walk and some analytical properties of the statistical integral combined with the mechanics of a constant function market maker (CFMM). We consider non-toxic or rather unspecific trading in this paper. Our main finding can be summarized in one sentence: For Brownian motion with a given volatility, IL and LVR have identical expectation values but vastly differing distribution functions. 
\end{abstract}
\section{Introduction}

Automated Market Makers (AMMs) are a fundamental innovation in decentralized finance (DeFi), facilitating the trading of digital assets without the need for a traditional order book. Instead, AMMs use liquidity pools and algorithms to determine prices and execute trades. AMMs rely on liquidity providers (LPers) to supply the assets that form the liquidity pools. Those LPers face the risk of impermanent loss (IL) or loss-vs-rebalancing (LVR), which occurs when the value of their deposited assets fluctuates compared to holding the assets separately. 

In this paper, we find that for small price movements, impermanent loss (IL) and loss-versus-rebalancing (LVR) yield identical results, despite their different interpretations and distributions. Specifically, while IL focuses solely on the difference between providing liquidity and holding assets over a given timeframe, LVR captures the cumulative effects of constant rebalancing. The paper is structured as follows: first, we introduce the basic setup and market maker model. Next, we analyze the dynamics of price fluctuations. In the following sections, we compare IL and LVR, and finally, we explore the impact of fees on liquidity provision.

\noindent{\it{Related literature:}}

 AMMs can be traced back to \cite{hanson2007logarithmic} and \cite{othman2013practical} with early implementations discussed in \cite{lehar2021decentralized}, \cite{capponi2021adoption}, and \cite{hasbrouck2022need}. Details of implementation are described in \cite{Adams20} and \cite{Adams21} as well as in a very recent textbook \cite{ottina2023automated}.

We study ways to optimize fees based on an arbitrage-only assumption. Uniswap v3 (\cite{Adams21}) addresses this problem by letting liquidity providers choose between different static fee tiers. Other automated market makers have implemented dynamic fees on individual pools, including Trader Joe v2.1 (\cite{mountainfarmer22joe}), Curve v2 (\cite{egorov21curvev2}) and Mooniswap (\cite{bukov20mooniswap}), Algebra (\cite{Volosnikov}), as well as \cite{Nezlobin2023}. Some of the general properties of toxic flow and loss versus rebalancing have been discussed in Refs.~\cite{Faycal1,Faycal2,Faycal3,milionis2024automated,crapis2023optimal,angeris2024multidimensional}

\section{The setup}\label{sec:setup}

\subsection{The automated market maker}

A constant function market maker (CFMM) with the formula \(xy - L^2 = 0\) describes the most basic automated market maker (AMM) model where \(x\) and \(y\) represent the quantities of two different tokens in a liquidity pool, and \(L\) is a constant that characterizes the pool's liquidity. In this model, the product of the quantities of the two tokens remains constant:

\[
xy = L^2
\]

This ensures that any trade which increases one token's amount \(x\) must decrease the other token's amount \(y\) and vice versa. The price of each token depends inversely on their respective quantities. Specifically, the price of token \(x\) in terms of token \(y\) is given by:

\[
p = \frac{y}{x}
\]

henceforth simply referred to as price. There is a relation between the token amounts and the price $p$ according to
\begin{eqnarray}
x(p)=\frac{L}{\sqrt{p}}  \quad {\rm{and}} \quad y(p)=L\sqrt{p}\;.
\end{eqnarray}

We assume from now on that there is a starting condition at time $t=0$ defined by
\begin{eqnarray}
x_0 y_0 =L^2
\end{eqnarray}
and price $p_0=y_0/x_0$. This allows to express the token numbers as a function of price according to
\begin{eqnarray}
x(p)=x_0\sqrt{\frac{p_0}{p}}  \quad {\rm{and}} \quad y(p)=x_0\sqrt{p_0 p}\;.
\end{eqnarray}

\subsection{Dynamical fluctuations of the price}

\begin{figure}[h]
    \centering
    \begin{minipage}{0.9\textwidth}
        \centering
        \includegraphics[width=\textwidth]{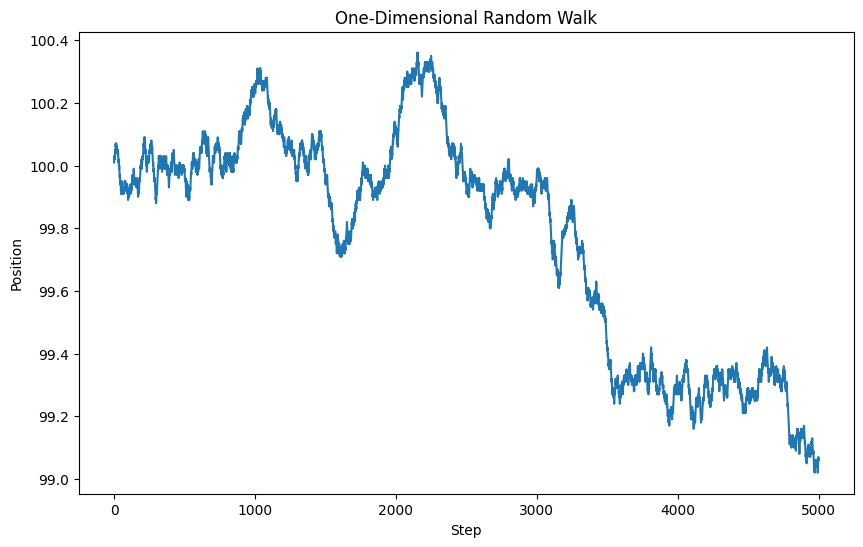}
        \caption{Price movement of a single run}
        \label{fig:mcprice}
    \end{minipage}
\end{figure}

\begin{figure}[h]
    \centering
    \begin{minipage}{0.8\textwidth}
        \centering
        \includegraphics[width=\textwidth]{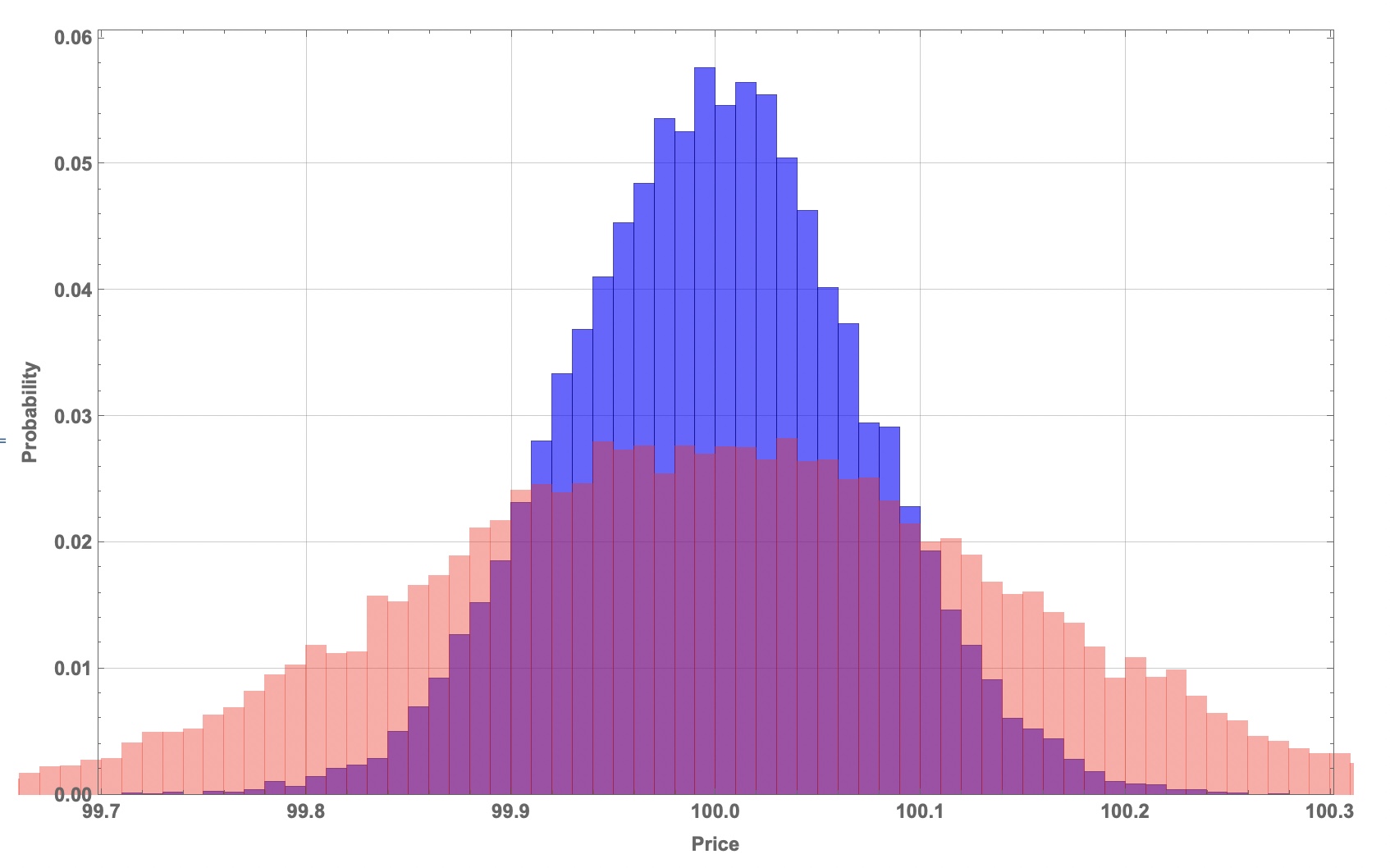}
        \caption{Distribution of final prices over 20000 runs}
        \label{fig:mcpricehistogram}
    \end{minipage}
\end{figure}
\begin{figure}[h]
    \centering
    \begin{minipage}{0.8\textwidth}
        \centering
        \includegraphics[width=\textwidth]{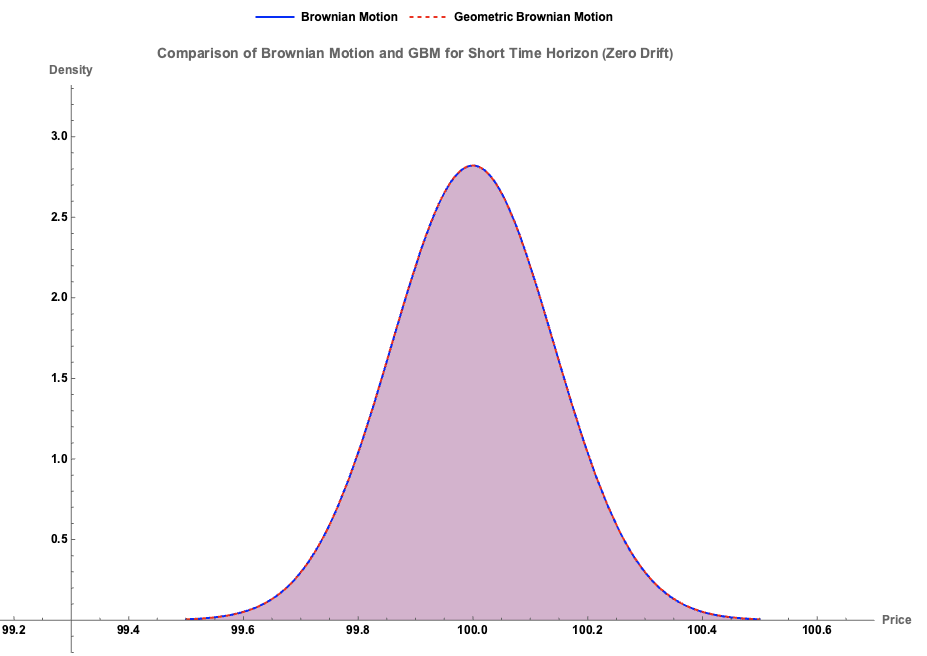}
        \caption{Comparison of the analytic distribution functions of Brownian motion and Geometric Brownian motion compared over $20000$ steps and for identical volatility. }
        \label{fig:mcpricehistogram}
    \end{minipage}
\end{figure}

For our analysis, we use a combination of numerical simulations and analytical calculations. For numerical simulations, we simulate the price movement with a simple random walk. We define it such that the price at time $t+1$ obtains from the price at time $t$ according to 
\begin{eqnarray}
p_{i+1}=p_i+\sigma_0\xi
\end{eqnarray}
where $\sigma_0$ is the variance and $\xi$ is $\pm 1$, both drawn with equal probability. We assume that the simulation starts at a price $p_0$. A characteristic single run for a starting price of $p_0=100$, $\sigma_0=0.001$, and $5000$ (blue) and $20000$ (orange) steps is shown in Fig.~\ref{fig:mcprice}. 

This leads to a price that follows Brownian motion. We have explicitly checked that for the simulation times we consider Brownian motion and Geometric Brownian motion are virtually indistinguishable, see Fig.~\ref{fig:brownian}. Therefore, we proceed with Brownian motion which has a simpler analytical limit. While the numerical formulation is useful for simulations, it is insightful to consider the distribution function. It is well known that the distribution of the price over time follows a Gauss distribution
\begin{eqnarray}
\rho (p,t)=\frac{1}{\sqrt{2\pi \sigma_0^2 t}}\exp{\left(-\frac{\left(p-p_0\right)^2}{2 \sigma_0^2 t}\right)}
\end{eqnarray}
where $p$ is the price, $p_0$ is the starting price, $\sigma_0$ is the volatility, and $t$ is time. This agrees with a histogram taken from $20000$ random walks, shown in Fig.~\ref{fig:mcpricehistogram}, again for $p_0=100$, $\sigma_0=0.01$, and $5000$ steps.

\section{IL vs LVR}

A common topic of discussion in the context of automated market makers (AMMs) is the relationship between impermanent loss (IL) and loss-versus-rebalancing (LVR), and which one serves as a more appropriate performance metric. In this section, we argue that for an infinitesimal price change from $p \to p + dp$ during a time interval $dt$, IL and LVR are mathematically identical. We will begin by reviewing the underlying mechanics of both metrics.

\subsection{Impermanent Loss (IL)}

Impermanent loss measures the difference between the value of a liquidity provider's (LP) position inside an AMM and the value the LP would have if they had simply held the assets outside the AMM (i.e., a HODL strategy). Before a price change, the value of the LP's position at price $p$ is given by
\[
V(p) = \frac{L}{\sqrt{p}} + \frac{1}{p}L\sqrt{p} = 2 \frac{L}{\sqrt{p}},
\]
where $L$ represents the initial liquidity provided by the LP.

After the price changes to $p + dp$, the value of the position becomes
\[
V(p + dp) = 2 \frac{L}{\sqrt{p + dp}} = 2 \frac{L}{\sqrt{p}} \sqrt{\frac{p}{p + dp}}.
\]
In contrast, if the LP had simply held the assets, the value of the position remains the same before the price change:
\[
\text{HODL}(p) = V(p),
\]
but after the price change, the HODL value is
\[
\text{HODL}(p + dp) = \frac{L}{\sqrt{p}} \left( 1 + \frac{p}{p + dp} \right).
\]

Thus, the impermanent loss after a small price change is given by
\[
IL(p, p + dp) = \frac{L}{\sqrt{p}} \left( 1 - \sqrt{\frac{p}{p + dp}} \right)^2.
\]

\subsection{Loss-Versus-Rebalancing (LVR)}

Loss-versus-rebalancing (LVR) addresses a different question. It examines how the value of a portfolio would change if, instead of being deposited in the AMM, the LP maintains a shadow portfolio that exactly mirrors the liquidity position over time.

After the price changes, the LP's position in the AMM adjusts such that the number of tokens $x$ and $y$ changes to
\[
x(p + dp) = \frac{L}{\sqrt{p}} \sqrt{\frac{p}{p + dp}}, \quad y(p + dp) = L \sqrt{p} \sqrt{\frac{p + dp}{p}}.
\]
To maintain a shadow portfolio that mimics the AMM position, the portfolio would need to rebalance by buying $\Delta y$ tokens:
\[
\Delta y = y(p + dp) - y(p) = L \sqrt{p} \left( \sqrt{\frac{p + dp}{p}} - 1 \right),
\]
at the price $p + dp$, which requires spending
\[
\Delta \bar{x} = \frac{L}{\sqrt{p}} \left( \sqrt{\frac{p}{p + dp}} - \frac{p}{p + dp} \right).
\]

The change in the LP’s token $x$ position in the AMM is
\[
\Delta x = \frac{L}{\sqrt{p}} \left( 1 - \sqrt{\frac{p}{p + dp}} \right),
\]
which turns out to be greater than the cost required to buy the additional $\Delta y$ on the open market. The savings in terms of rebalancing, or the LVR, is therefore
\[
\Delta LVR(p, p + dp) = \frac{L}{\sqrt{p}} \left( 1 - \sqrt{\frac{p}{p + dp}} \right)^2.
\]

\subsection{Comparing IL and LVR}

From the above, it is clear that for an infinitesimal price change from $p$ to $p + dp$, both impermanent loss and loss-versus-rebalancing give the same result:
\[
\frac{L}{\sqrt{p}} \left( 1 - \sqrt{\frac{p}{p + dp}} \right)^2.
\]
This shows that for small price movements, IL and LVR are mathematically equivalent. This is also intuitive since they are just two different points of view on the same thing.

\subsection{Key Difference Between IL and LVR}

While IL and LVR are identical for small price changes, they differ in their broader interpretation, especially if we consider a position over an extended time frame. IL focuses solely on the difference between providing liquidity and holding the assets at specific price points, the end points, without considering the price path between the points. On the other hand, LVR aggregates the infinitesimal changes that occur as the price evolves over time, capturing the cumulative effect of constant rebalancing.

\subsection{Differential equation for LVR}

While IL only cares about the start and end points of the price path, LVR is summed up along the whole path. To better describes this, we now convert the expression for LVR into a differential equation. We start with assuming that the changes in price $dp \ll p$ and expand the expression for LVR according to 
\[
\Delta LVR(p, p + dp) = \frac{L}{\sqrt{p}} \left( 1 - \sqrt{\frac{p}{p + dp}} \right)^2 \approx \frac{L}{4} \frac{dp^2}{p^{5/2}}
\]

For Brownian motion it is well known that $d p^2=\sigma_0 d t$ meaning we 
find
\begin{eqnarray}
\Delta{\rm{LVR}}(p,\Delta t)=\frac{L}{4} \frac{\sigma_0^2 dt}{p^{5/2}}.
\end{eqnarray}
Performing the limit $d t \to 0$ we can convert this into a differential equation according to
\begin{eqnarray}
\frac{d {\rm{LVR}(p(t))}}{dt}=L\frac{\sigma_0^2}{4 p(t)^{5/2}}\;.
\end{eqnarray}
We note that in this differential equation the time dependence is implicit in the trajectory $p(t)$ and therefore depends on the individual realization.

To summarize: IL is calculated between start and end points of an observed time frame and as such does not care about intermediate losses. LVR, on the other hand is updated after every price change on the trajectory. They are connected in the following sense: LVR sums up IL of price changes within the individual time unit $\d t$. A naive expectation is that LVR should be much bigger than IL because one is summing up pieces all the time. We will find that this expectation is wrong on average but true for most paths.

We will devote the following section to finding a better understanding of their relation. 

\section{Analysis of IL and LVR}

We use two tools in this section: numerical simulations based on the random walk as well as statistical properties of the Gaussian distribution. We find, as expected, excellent agreement between the two.

\subsection{Random walk analysis}

\begin{figure}[h]
    \centering
    \includegraphics[width=0.8\textwidth]{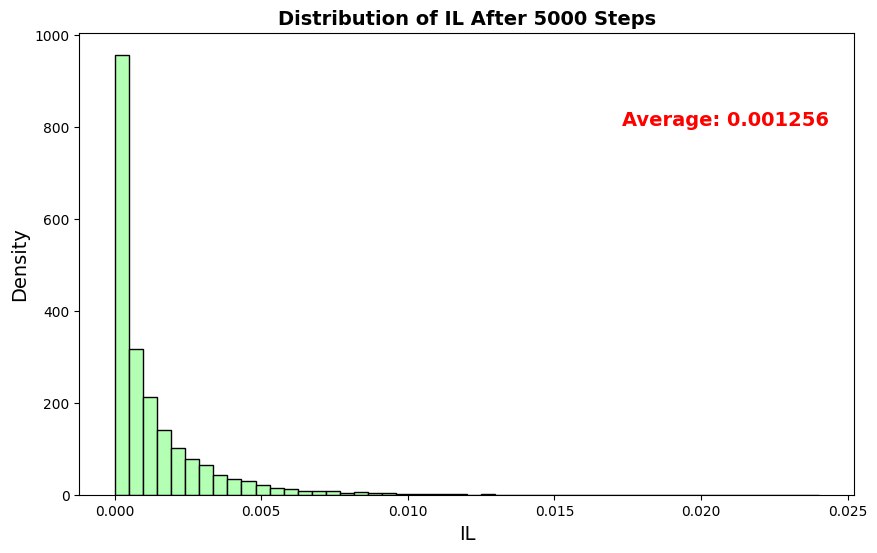}
    \caption{Distribution of IL over 20000 runs.}
    \label{fig:MCILvsLVRa}
\end{figure}

\begin{figure}[h]
    \centering
    \includegraphics[width=0.8\textwidth]{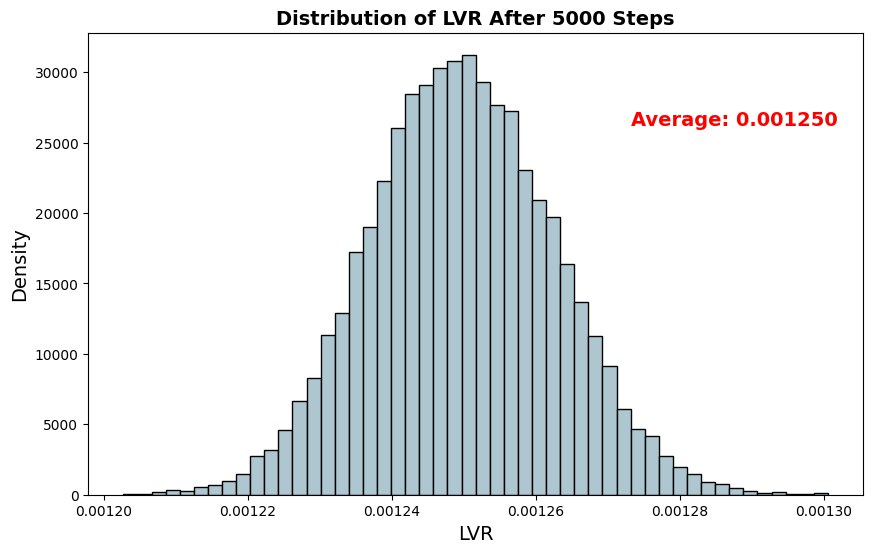}
    \caption{Distribution of IL over 20000 runs.}
    \label{fig:MCILvsLVRb}
\end{figure}
In all the subsequent plots we have chosen the following setting. We choose a starting price of $p_0=100$ and $x_0=100$, as well as $\sigma_0=0.01$. A single run consists of an evolution of $5000$ time steps and we perform 20000 runs. We record both histograms of the runs as well as averages. 

We calculated IL (Fig.~\ref{fig:MCILvsLVRa}) and LVR (Fig.~\ref{fig:MCILvsLVRb}) for the same settings. We find, maybe surprisingly at this point, that while both quantities have very different distribution functions, they possess the same average. This has been verified for many other parameter settings so this is not a coincidence but independent of parameters. 

The shape of the distribution functions can easily be rationalized. IL only measures a loss at the end point relative to the starting point. Most trajectories for the random walk end in a position relatively close to $p_0$. Those trajectories contribute very little IL meaning the majority of trajectories has sub-average IL. LVR, on the other hand, realizes a loss at every step. Since the trajectories predominantly hover around $p_0$, those losses are roughly the same for every trajectory at every time step. Consequently, most trajectories collect average LVR. The surprising insight is that the average IL and the average LVR agree within statistical accuracy. An immediate question is whether one is a more useful metric to quantify losses than the other. The advantage of LVR is that looking at a number of positions gives a good chance to identify the correct value while with IL the bulk of the contributions from trajectories with little probability so IL will easily underestimate the actual loss (if more positions with different starting points were considered). 

We will now use the properties of the Gaussian distribution to show the agreement between the averages is no coincidence.

\subsection{Analytical treatment}

At this point we cannot refrain from stating that the following procedure has a very prominent counterpart in the theory of quantum mechanics, which is the Feynman path integral. The Feynamn path integral sums up all the possible paths that a particle could take to go from one place in space-time to another. If we replace space with price, we have the correspondence (since this ia a classical problem it is in fact more similar to the Wiener integral).

\begin{figure}
\centering
\includegraphics[width=0.8\textwidth]{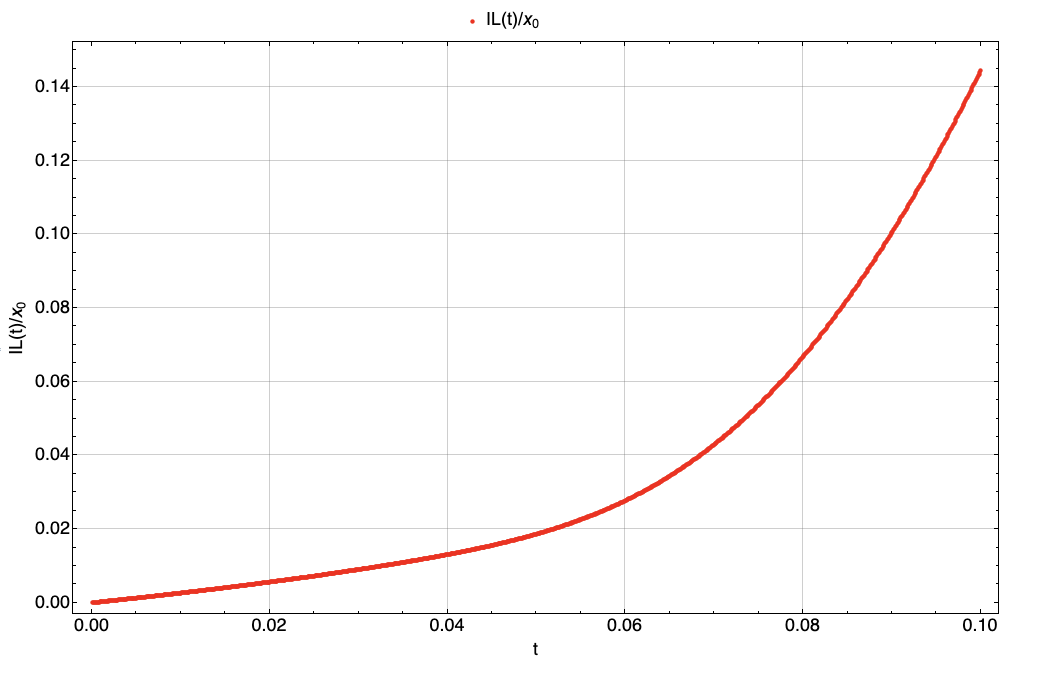}\caption{Expected IL as a function of time, $\langle IL (t) \rangle$, measured in units of $x_0$. For this plot we chose $\sigma_0=p_0=1$.}\label{fig:IL}
\end{figure}

In a first step, we analyze the expected IL as a function of time. It turns out that this quantity can readily be calculated from summing IL over all possible paths in price space:
\begin{eqnarray}
\langle {\rm{IL}}(t) \rangle &=& \int dp\frac{{\rm{IL}}(p)}{\sqrt{2\pi \sigma_0^2 t}}\exp{\left(-\frac{\left(p-p_0\right)^2}{2 \sigma_0^2 t}\right)} \nonumber \\ &=& \frac{x_0}{\sqrt{2\pi \sigma_0^2 t}}\int dp \left(1-\sqrt{\frac{p_0}{p}} \right)^2\exp{\left(-\frac{\left(p-p_0\right)^2}{2 \sigma_0^2 t}\right)} \nonumber \\ &=& \frac{x_0}{\sqrt{\pi}}\int dp \left(1-\sqrt{\frac{p_0}{p_0+\sqrt{2\sigma_0^2t}p}} \right)^2\exp{\left(-p^2\right)}\;.
\end{eqnarray}
We find that this integral does not extend to $-\infty$ but has to be cut off at $-p_0/\sqrt{2 \sigma_0^2 t}$. For practical purposes and short times $t\ll 2\sigma_0^2p_0^2$ we can expand the integrand to yield
\begin{eqnarray}
\langle {\rm{IL}}(t) \rangle \approx  \frac{x_0 \sigma_0^2}{4 p_0^2} t \int dp \frac{2 p^2 }{\sqrt{\pi}}\exp{\left(-p^2\right)}=\frac{x_0 \sigma_0^2}{4 p_0^2} t
\end{eqnarray}
which implies that the expected IL increases linearly with time (this integral can now be extended all the way to $-\infty$). A full numerical solution of the integral with its actual boundaries is shown in Fig.~\ref{fig:IL} but not important for our discussion. It just serves as a proof of validity of our expansion. 

The analytical formula captures the linear part well and could also be used to characterize the deviation if higher orders were takein into account. 

We now move to average LVR. It can be calculated from

\begin{eqnarray}
\langle {\rm{LVR}}(t) \rangle &=& \int dp \int_0^t dt' \, \rho(p,t') \frac{d {\rm{LVR}}(p)}{dt'} \nonumber \\ 
&=& \frac{x_0 \sigma_0 \sqrt{p_0}}{4\sqrt{2\pi }} \int dp \int_0^t dt' \frac{1}{\sqrt{t'}p^{5/2}}\exp{\left(-\frac{\left(p-p_0\right)^2}{2 \sigma_0^2 t'}\right)} \nonumber \\ 
&=& \frac{x_0 \sigma_0^2 \sqrt{p_0}}{4\sqrt{\pi }} \int dp \int_0^t dt' \frac{1}{(p_0+\sqrt{2\sigma_0^2 t'}p)^{5/2}}\exp{\left(-p^2\right)} \;.
\end{eqnarray}
We can expand the integrand to lowest order as before and get
\begin{eqnarray}
\langle {\rm{LVR}}(t) \rangle &\approx&  \frac{x_0 \sigma_0^2 }{4\sqrt{\pi }p_0^2} \int dp \int_0^t dt' \exp{\left(-p^2\right)} =\frac{x_0 \sigma_0^2 }{4p_0^2}t\;.
\end{eqnarray}
We thus conclude that we manged to show that $\langle {\rm{IL}}(t) \rangle=\langle {\rm{LVR}}(t) \rangle$, as we already observed from the numerical simulation. Furthermore, the numerical findings are in excellent agreement with the analytical predictions. The analytical prediction for the plots shown in Fig.~\ref{fig:MCILvsLVRa} and Fig.~\ref{fig:MCILvsLVRb} is $\langle {\rm{LVR}}(5000) \rangle=\langle {\rm{IL}}(5000) \rangle=0.00125$ for both averages.

\section{Conclusion}

In conclusion, this paper demonstrates that while impermanent loss (IL) and loss-versus-rebalancing (LVR) are distinct in their interpretations, they are mathematically equivalent for small price movements. Through both numerical simulations and analytical calculations, we show that, {\it{on average}}, IL and LVR behave in exactly the same way, even though their distributions differ significantly. Under stable market conditions, we find that LVR of the past has a good chance of having predictive power for the future expected LVR and even IL. Overall, this suggests that LVR, with its ability to capture intermediate price changes, may serve as a more reliable metric for evaluating and predicting liquidity provider performance. However, looking at the IL distribution function reveals one feature: the probability of incurring IL that is below the expectation value (and that of LVR) is quite high and there is a reasonable hope that LVR provides a worse-than-real scenario when P$\&$L is considered. We will discuss the dynamics of fees and actual arbitrage dynamics in an upcoming paper where we also introduce a novel dynamical fee algorithm that reduces arbitrage incurred LVR by more than $30\%$ compared to currently employed fixed fee tiers like employed for instance in Uniswap v3 pools.

\section*{Acknowledgments}

We would like to express our gratitude to Ciamac Moallemi, Guillaume Lambert, Gigasafu, and Atis Elsts for discussions and insightful feedback.

{\small
  \bibliographystyle{ACM-Reference-Format}
  \bibliography{references}


\begin{thebibliography}{19}


\ifx \showCODEN    \undefined \def \showCODEN     #1{\unskip}     \fi
\ifx \showDOI      \undefined \def \showDOI       #1{#1}\fi
\ifx \showISBNx    \undefined \def \showISBNx     #1{\unskip}     \fi
\ifx \showISBNxiii \undefined \def \showISBNxiii  #1{\unskip}     \fi
\ifx \showISSN     \undefined \def \showISSN      #1{\unskip}     \fi
\ifx \showLCCN     \undefined \def \showLCCN      #1{\unskip}     \fi
\ifx \shownote     \undefined \def \shownote      #1{#1}          \fi
\ifx \showarticletitle \undefined \def \showarticletitle #1{#1}   \fi
\ifx \showURL      \undefined \def \showURL       {\relax}        \fi
\providecommand\bibfield[2]{#2}
\providecommand\bibinfo[2]{#2}
\providecommand\natexlab[1]{#1}
\providecommand\showeprint[2][]{arXiv:#2}

\bibitem[Adams et~al\mbox{.}(2020)]%
        {Adams20}
\bibfield{author}{\bibinfo{person}{Hayden Adams}, \bibinfo{person}{Noah Zinsmeister}, {and} \bibinfo{person}{Dan Robinson}.} \bibinfo{year}{2020}\natexlab{}.
\newblock \bibinfo{booktitle}{\emph{Uniswap v2 Core}}.
\newblock
\urldef\tempurl%
\url{https://uniswap.org/whitepaper.pdf}
\showURL{%
Retrieved Jun 12, 2023 from \tempurl}


\bibitem[Adams et~al\mbox{.}(2021)]%
        {Adams21}
\bibfield{author}{\bibinfo{person}{Hayden Adams}, \bibinfo{person}{Noah Zinsmeister}, \bibinfo{person}{Moody Salem}, \bibinfo{person}{River Keefer}, {and} \bibinfo{person}{Dan Robinson}.} \bibinfo{year}{2021}\natexlab{}.
\newblock \bibinfo{booktitle}{\emph{Uniswap v3 Core}}.
\newblock
\urldef\tempurl%
\url{https://uniswap.org/whitepaper-v3.pdf}
\showURL{%
Retrieved Jun 12, 2023 from \tempurl}


\bibitem[Angeris et~al\mbox{.}(2024)]%
        {angeris2024multidimensional}
\bibfield{author}{\bibinfo{person}{Guillermo Angeris}, \bibinfo{person}{Theo Diamandis}, {and} \bibinfo{person}{Ciamac Moallemi}.} \bibinfo{year}{2024}\natexlab{}.
\newblock \bibinfo{title}{Multidimensional Blockchain Fees are (Essentially) Optimal}.
\newblock
\newblock
\showeprint[arxiv]{2402.08661}~[cs.GT]


\bibitem[Bukov and Melnik(2020)]%
        {bukov20mooniswap}
\bibfield{author}{\bibinfo{person}{Anton Bukov} {and} \bibinfo{person}{Mikhail Melnik}.} \bibinfo{year}{2020}\natexlab{}.
\newblock \bibinfo{booktitle}{\emph{Mooniswap by 1inch.exchange}}.
\newblock
\urldef\tempurl%
\url{https://mooniswap.exchange/docs/MooniswapWhitePaper-v1.0.pdf}
\showURL{%
Retrieved Sept 18, 2023 from \tempurl}


\bibitem[Capponi and Jia(2021)]%
        {capponi2021adoption}
\bibfield{author}{\bibinfo{person}{Agostino Capponi} {and} \bibinfo{person}{Ruizhe Jia}.} \bibinfo{year}{2021}\natexlab{}.
\newblock \showarticletitle{The adoption of blockchain-based decentralized exchanges}.
\newblock \bibinfo{journal}{\emph{arXiv preprint arXiv:2103.08842}} (\bibinfo{year}{2021}).
\newblock


\bibitem[Cartea et~al\mbox{.}(2022)]%
        {Faycal2}
\bibfield{author}{\bibinfo{person}{{\'A}lvaro Cartea}, \bibinfo{person}{Fay{\c{c}}al Drissi}, {and} \bibinfo{person}{Marcello Monga}.} \bibinfo{year}{2022}\natexlab{}.
\newblock \showarticletitle{Decentralised Finance and Automated Market Making: Predictable Loss and Optimal Liquidity Provision}.
\newblock  (\bibinfo{date}{November 10} \bibinfo{year}{2022}).
\newblock
\newblock
\shownote{Available at SSRN: \url{https://ssrn.com/abstract=4273989} or \url{http://dx.doi.org/10.2139/ssrn.4273989}}.


\bibitem[Cartea et~al\mbox{.}(2023)]%
        {Faycal3}
\bibfield{author}{\bibinfo{person}{{\'A}lvaro Cartea}, \bibinfo{person}{Fay{\c{c}}al Drissi}, \bibinfo{person}{Leandro S{\'a}nchez-Betancourt}, \bibinfo{person}{David Siska}, {and} \bibinfo{person}{Lukasz Szpruch}.} \bibinfo{year}{2023}\natexlab{}.
\newblock \showarticletitle{Automated Market Makers Designs Beyond Constant Functions}.
\newblock  (\bibinfo{date}{May 25} \bibinfo{year}{2023}).
\newblock
\newblock
\shownote{Available at SSRN: \url{https://ssrn.com/abstract=4459177} or \url{http://dx.doi.org/10.2139/ssrn.4459177}}.


\bibitem[Crapis et~al\mbox{.}(2023)]%
        {crapis2023optimal}
\bibfield{author}{\bibinfo{person}{Davide Crapis}, \bibinfo{person}{Ciamac~C. Moallemi}, {and} \bibinfo{person}{Shouqiao Wang}.} \bibinfo{year}{2023}\natexlab{}.
\newblock \bibinfo{title}{Optimal Dynamic Fees for Blockchain Resources}.
\newblock
\newblock
\showeprint[arxiv]{2309.12735}~[cs.GT]


\bibitem[Egorov and GmbH)(2021)]%
        {egorov21curvev2}
\bibfield{author}{\bibinfo{person}{Michael Egorov} {and} \bibinfo{person}{Curve Finance (Swiss~Stake GmbH)}.} \bibinfo{year}{2021}\natexlab{}.
\newblock \bibinfo{booktitle}{\emph{Automatic market-making with dynamic peg}}.
\newblock
\urldef\tempurl%
\url{https://classic.curve.fi/files/crypto-pools-paper.pdf}
\showURL{%
Retrieved Sept 18, 2023 from \tempurl}


\bibitem[Hanson(2007)]%
        {hanson2007logarithmic}
\bibfield{author}{\bibinfo{person}{Robin Hanson}.} \bibinfo{year}{2007}\natexlab{}.
\newblock \showarticletitle{Logarithmic markets coring rules for modular combinatorial information aggregation}.
\newblock \bibinfo{journal}{\emph{The Journal of Prediction Markets}} \bibinfo{volume}{1}, \bibinfo{number}{1} (\bibinfo{year}{2007}), \bibinfo{pages}{3--15}.
\newblock


\bibitem[Hasbrouck et~al\mbox{.}(2022)]%
        {hasbrouck2022need}
\bibfield{author}{\bibinfo{person}{Joel Hasbrouck}, \bibinfo{person}{Thomas~J Rivera}, {and} \bibinfo{person}{Fahad Saleh}.} \bibinfo{year}{2022}\natexlab{}.
\newblock \showarticletitle{The need for fees at a dex: How increases in fees can increase dex trading volume}.
\newblock \bibinfo{journal}{\emph{Available at SSRN}} (\bibinfo{year}{2022}).
\newblock


\bibitem[Lehar and Parlour(2021)]%
        {lehar2021decentralized}
\bibfield{author}{\bibinfo{person}{Alfred Lehar} {and} \bibinfo{person}{Christine~A Parlour}.} \bibinfo{year}{2021}\natexlab{}.
\newblock \showarticletitle{Decentralized exchanges}.
\newblock \bibinfo{journal}{\emph{Available at SSRN 3905316}} (\bibinfo{year}{2021}).
\newblock


\bibitem[Milionis et~al\mbox{.}(2024)]%
        {milionis2024automated}
\bibfield{author}{\bibinfo{person}{Jason Milionis}, \bibinfo{person}{Ciamac~C. Moallemi}, \bibinfo{person}{Tim Roughgarden}, {and} \bibinfo{person}{Anthony~Lee Zhang}.} \bibinfo{year}{2024}\natexlab{}.
\newblock \bibinfo{title}{Automated Market Making and Loss-Versus-Rebalancing}.
\newblock
\newblock
\showeprint[arxiv]{2208.06046}~[q-fin.MF]


\bibitem[MountainFarmer et~al\mbox{.}(2022)]%
        {mountainfarmer22joe}
\bibfield{author}{\bibinfo{person}{MountainFarmer}, \bibinfo{person}{Louis}, \bibinfo{person}{Hanzo}, \bibinfo{person}{Wawa}, \bibinfo{person}{Murloc}, {and} \bibinfo{person}{Fish}.} \bibinfo{year}{2022}\natexlab{}.
\newblock \bibinfo{booktitle}{\emph{JOE v2.1 Liquidity Book}}.
\newblock
\urldef\tempurl%
\url{https://github.com/traderjoe-xyz/LB-Whitepaper/blob/main/Joe%20v2%20Liquidity%20Book%20Whitepaper.pdf}
\showURL{%
Retrieved Sept 18, 2023 from \tempurl}


\bibitem[Nezlobin(2023)]%
        {Nezlobin2023}
\bibfield{author}{\bibinfo{person}{Alex Nezlobin}.} \bibinfo{year}{2023}\natexlab{}.
\newblock \bibinfo{booktitle}{\emph{Twitter thread}}.
\newblock
\urldef\tempurl%
\url{https://twitter.com/0x94305/status/1674857993740111872}
\showURL{%
Retrieved Dec 3, 2023 from \tempurl}


\bibitem[Othman et~al\mbox{.}(2013)]%
        {othman2013practical}
\bibfield{author}{\bibinfo{person}{Abraham Othman}, \bibinfo{person}{David~M Pennock}, \bibinfo{person}{Daniel~M Reeves}, {and} \bibinfo{person}{Tuomas Sandholm}.} \bibinfo{year}{2013}\natexlab{}.
\newblock \showarticletitle{A practical liquidity-sensitive automated market maker}.
\newblock \bibinfo{journal}{\emph{ACM Transactions on Economics and Computation (TEAC)}} \bibinfo{volume}{1}, \bibinfo{number}{3} (\bibinfo{year}{2013}), \bibinfo{pages}{1--25}.
\newblock


\bibitem[Ottina et~al\mbox{.}(2023)]%
        {ottina2023automated}
\bibfield{author}{\bibinfo{person}{M. Ottina}, \bibinfo{person}{P.J. Steffensen}, {and} \bibinfo{person}{J. Kristensen}.} \bibinfo{year}{2023}\natexlab{}.
\newblock \bibinfo{booktitle}{\emph{Automated Market Makers: A Practical Guide to Decentralized Exchanges and Cryptocurrency Trading}}.
\newblock \bibinfo{publisher}{Apress}.
\newblock
\showISBNx{9781484286159}
\urldef\tempurl%
\url{https://books.google.nl/books?id=BKU6zwEACAAJ}
\showURL{%
\tempurl}


\bibitem[Volosnikov et~al\mbox{.}(2022)]%
        {Volosnikov}
\bibfield{author}{\bibinfo{person}{Vladislav Volosnikov}, \bibinfo{person}{Vladimir Tikhomirov}, \bibinfo{person}{Ilya Azhel}, {and} \bibinfo{person}{Ilya Chizh}.} \bibinfo{year}{2022}\natexlab{}.
\newblock \bibinfo{booktitle}{\emph{ALGEBRA Ecosystem: Decentralized exchange}}.
\newblock
\urldef\tempurl%
\url{https://algebra.finance/static/Algerbra%20Tech%20Paper-15411d15f8653a81d5f7f574bfe655ad.pdf}
\showURL{%
\tempurl}


\bibitem[Álvaro Cartea and Monga(2023)]%
        {Faycal1}
\bibfield{author}{\bibinfo{person}{Fayçal~Drissi Álvaro Cartea} {and} \bibinfo{person}{Marcello Monga}.} \bibinfo{year}{2023}\natexlab{}.
\newblock \showarticletitle{Predictable Losses of Liquidity Provision in Constant Function Markets and Concentrated Liquidity Markets}.
\newblock \bibinfo{journal}{\emph{Applied Mathematical Finance}} \bibinfo{volume}{30}, \bibinfo{number}{2} (\bibinfo{year}{2023}), \bibinfo{pages}{69--93}.
\newblock
\urldef\tempurl%
\url{https://doi.org/10.1080/1350486X.2023.2277957}
\showDOI{\tempurl}
\showeprint{https://doi.org/10.1080/1350486X.2023.2277957}


\end{thebibliography}
}

\end{document}